# A Computational Phase Field Study of Conducting Channel Formation in Dielectric Thin Films: A View Towards the Physical Origins of Resistive Switching

John F. Sevic and Nobuhiko P. Kobayashi

Department of Electrical Engineering, Baskin School of Engineering, University of California, Santa Cruz, California 95064 USA

March 2, 2019

*Abstract:* *A phase field method is used to computationally study conducting channel morphology of resistive switching thin film structures. Our approach successfully predicts the formation of conducting channels in typical dielectric thin film structures, comparable to a range of resistive switches, offering an alternative computational formulation based on metastable states treated at the atomic scale. In contrast to previous resistive switching thin film models, our formulation makes no a priori assumptions on conducting channel morphology and its fundamental transport mechanisms.*

## I. INTRODUCTION

Physical and chemical properties of dielectric thin films for resistive switching are governed by the dynamical behavior of intrinsic atomic metastable states of thin film materials and the presence of irreversible states endowed by specific preparation processes. Atomic metastable states may arise from various phenomena, such as Mott localization or change in configurational entropy, producing such effects as insulator-metal transition and hysteresis often seen in appropriately prepared dielectric thin films species[1,2,3,4]. When two chemically distinct and spatially distinct materials separated by an interface are forced to interact, they would, in the steady-state, produce either a uniform solid solution or decompose into two distinct phases via phase separation, depending on their mutual affinity associated with their unique bulk free-energy density. However, the morphological evolution of the interface separating each material is inherently dynamic and naturally produces complex microstructure resulting from various interactions of bulk free-energy, interface energy, electrothermal phenomena, and electrochemical phenomena the interface experiences[5]. Therefore, distinctive microstructure is expected to appear at the interface to self-consistently minimize the total energy, which in turn result in characteristic transport properties fundamental to, for instance, resistive switching behavior exhibited by dielectric thin films [4,6].

Various qualitative models have been adopted to study resistive switching behavior of dielectric thin films, largely based on formation of conducting channels composed of clusters of charged species[7,8,9,10]. These models invoke both electronic transport and ionic transport, often treated as electrothermal and electrochemical processes, respectively, reproducing both unipolar and bipolar resistive switching behavior. This conducting channel formalism suggests an initial irreversible growth of cluster-like aggregates composed of charged species forming conducting channels, as illustrated by Figure 1. These conducting channels subsequently form and rupture under the influence of an external electric potential, yielding resistive switching behavior.

Numerous computational studies have adopted the conducting channel formalism to advance theories on resistive switching that occurs in thin films made of dielectric materials. Xu posed a qualatative conducting channel model based on experimental data suggesting clustering and bifurcation of thread-like conducting channels to explain set and reset processes of bipolar resistive RAM (RRAM) devices[7]. Pan presented a similar study, with empirical evidence supporting formation and rupture of conducting channels corresponding to low-resistance and high-resistance states, respectively, of their ZnO memristor[8]. Ielmini presented a self-consistent electrothermal computational study based on a thermally activated ion migration model demonstrating conducting channel formation and rupture[9]. Using an analytical formulation, the approach correctly resolved set and reset behavior of a generic bipolar RRAM device by dynamic coupling of the channel resistance, charge transport and temperature, providing insight on conducting channel dynamic behavior. Gibson demonstrated steady-state negative differential resistance in niobium oxide selectors using a compact behavioral model self-consistently coupled to an electrothermal network. Their model demonstrated insulator-metal transition is not required for the presence of negative differential resistance, being due instead to Frenkel transport[11]. In contrast, Sevic posed a continuum-based computational study of the dynamical evolution of niobium oxide selector electroforming[12,13].

The conducting channel formalism has been complimented by several empirical studies on channel morphology and effect on transport properties of mobile charge carriers as the mechanism fundamental to resistive switching behavior. Yang and Strachan, in two complimentary studies of titanium oxide thin films using atomic force microscopy and transmission electron microscopy, provided experimental evidence of regions of dynamic conductivity modulation corresponding to set and reset. Their studies also indicated resistive switching behavior of their structure is related to a localized partial reduction of titanium dioxide and subsequent formation of a metallic conducting channel[14,15]. Miao studied tantalum oxide thin films using pressure-modulated conductance microscopy to identify regions of dynamic conductivity modulation. Transmission electron crosssectioning of these modulated regions suggested presence of conductive channels[16]. Ahmed used electron energy loss spectroscopy to observe formation and rupture of the oxygen deficient conducting channels goverening charge transport phenomema in their perovskite strontium titanate metal-insulator-metal (MIM) structure[10]. Their results suggest the presence of complex interface microstructure and spinodal decomposition distinguishing the semi-metallic conducting channels embedded in amorphous bulk.

While the conducting channel formalism qualitatively explains resistive switching behavior of dielectric thin films, in contrast, previous computational studies have largely adopted a continuum transport formulation. Selfconsistent solution of the continuum transport equations dynamically emulates advection and diffusion of thermally-activated charged species, and their interaction with local electric potential and temperature, to model bulk resistive switching phenomena. The continuum formulation, however, depends vitally on an *a priori* conducting channel transport model and correct identification of diffusion and mobility expressions for each specific transport mechanism, for example Frenkel transport for ionic vacancy conduction[11, 12, 13, 17, 18, 19, 20, 21].

A computational formulation that does not *a priori* impose assumptions on conducting channel morphology, transport phenomena, or interface uniformity, and instead treats resistive switching from its origin at the atomic-scale, may offer significant advantages over existing methods. Such a method might model the dynamical evolution of cluster-like charged aggregates, as illustrated by Figure 1, subject to their atomic and interfacial electrothermal interaction, naturally producing conducting channels in a non-conducting host. The phase field method is one such method[22, 23, 24, 25].

In this paper, we apply a phase field method to study dynamical evolution of conducting channels that influence resistive switching behavior exhibited by dielectric thin films. With the phase field method, the assumptions of an *a priori* conducting channel model and the presence of specific transport phenomena to explain resistive switching are abandoned, and our model is instead formulated as a diffuse interface problem subject to a variational principle[26]. Our approach successfully predicts the formation of conducting channels in typical dielectric thin film structures comparable to a range of resistive switches, offering an alternative computational formulation based on metastable states treated at the atomic scale, requiring no assumptions on conducting channel morphology and its fundamental transport mechanisms. Further, our approach applies to both electronic transport and ionic transport, *e.g.* ionic oxygen vacancies, however the current study focuses exclusively on electronic transport of dielectric thin films and unipolar resistive switching.

## II. SELF-CONSISTENT PHASE FIELD FORMULATION

Charge transport properties of thin films, governed by the dynamical behavior of atomic metastable states, are naturally treated by the phase field method. In contrast to molecular dynamics, which tracks the motion of each charge carrier, the phase field formulation tracks the dynamical evolution of the envelope of clusters of charge carriers whose aggregate boundary, an a priori unknown, forms a conducting channel interface within the non-conducting host, as illustrated by Figure 1. The phase field formulation thus avoids the mathematically onerous problem of expressing dynamic boundary conditions over an interface whose location is part of the unknown solution. In our study, self-consistent solution of the phase field equation yields the dynamical evolution of the interface formed between the conducting state and non-conducting state, both of which co-exist in a dielectric thin film, subject to local conservation laws.

To apply the phase field formulation, consider Figure 2 illustrating a pristine dielectric thin film structure composed of a conducting region and a non-conducting region, separated by an interface represented by the dotted horizontal line. This initial structure is viewed as an as-fabricated resistive switch comprising a dielectric thin film in which two distinct regions are separated by an interface, comprising a resistive switch made of a dielectric thin film in which mobile charges are initially distributed in a certain way creating the two regions, one conducting and the other non-conducting. This specific initial structure allows a double-well free-energy density function and the diffuse interface approximation to suitably describe dynamical structural evolution within our dielectric thin films. The bulk free-energy density function associated with the dielectric thin film structure of Figure 2 is given by

$$f_{bulk}(c) = A \times \left[c(\vec{r},t) - c_1\right]^2 \left[c(\vec{r},t) - c_2\right]^2 \quad (1)$$

where $A$ is the magnitude of the double-well potential, $c_1$ and $c_2$ represent normalized concentration of the conducting and non-conducting states, respectively, and $c(r,t)$ is the concentration variable, an unknown[27]. The concentration is bounded to the interval $0 \leq c(r,t) \leq 1$, with unity corresponding to the pure conducting state and zero corresponding to the pure non-conducting state. Here $r$ represents a location in the $(x,y)$-plane of the structure of Figure 2 and $t$ is time.

Interaction of the bulk free-energy and interface energy with electric potential externally applied to the structure of Figure 2 is modeled by the following electrostatic energy term

$$g_{elec}(c, V) = \frac{q}{\Omega} V(\vec{r},t) c(\vec{r},t) \quad (2)$$

where $V(r, t)$ is electric potential between the top and bottom edges of the structure, $q$ is electronic charge, and $\Omega$ is a differential volume unit[28]. The free-energy functional obtains by combining Equations 1 and 2 with an interface gradient energy term to yield

$$F = \int_R \left[ f_{bulk}(c) + \frac{\kappa}{2}\nabla^2 c(\vec{r},t) + g_{elec}(c,V) \right] d\vec{r} \quad (3)$$

where $\kappa$ is an interfacial gradient energy term that relates to energy stored per unit application of the potential of the gradient of $c(r,t)$ and the integration is over $R$, the entire thin film structure of Figure 2. Interfacial gradient energy is assumed to be uniformly constant along the interface.

The phase field transport equation is found by seeking an energy-minimizing stationary state of the free-energy, identified by finding the first-order variation of free-energy functional Equation 3. From the first-order variation of free-energy functional Equation 3, the phase field flux obtains

$$J_{PF} = M\nabla \left[ \frac{\partial f_{bulk}(c)}{\partial c} - \nabla \cdot \kappa \nabla c(\vec{r},t) - \frac{q}{\Omega}V(\vec{r},t) \right] \quad (4)$$

where $M$ is an associated mobility of the phase field flux in phase-space, and assumed constant. Since the concentration variable, $c(r,t)$, is conserved, the phase field conservation law obtains

$$\frac{\partial c(\vec{r},t)}{\partial t} = \nabla \cdot J_{PF} \quad (5)$$

This is the Cahn-Hilliard phase field equation in concentration variable $c(r,t)$ for our dielectric thin film model[29]. The dynamical evolution of the conducting channel, created at the interface formed by the conducting and non-conducting states, under the influence of an externally applied electric potential, is described by phase field conservation Equation 5 self-consistently coupled to the electronic Laplace equation[13, 26]. This yields

$$\frac{\partial c(\vec{r},t)}{\partial t} = \nabla \cdot M\nabla \left[ \frac{\partial f_{bulk}(c)}{\partial c} - \nabla \cdot \kappa \nabla c - \frac{q}{\Omega}V \right] \quad (6a)$$

$$\nabla \cdot \sigma\left[c(\vec{r},t)\right] \nabla V(\vec{r},t) = 0 \quad (6b)$$

where local conductivity, $\sigma[c(r,t)]$, is a linear positive-monotonic function of concentration, $c(r,t)$[30]. For the current computational study, an isothermal assumption is made for the initial forming process. Our current focus is to assess the applicability of the phase field method in the context of conducting channel formation in dielectric thin films, although we are fully aware of the importance of thermal effects used in our study and their intrinsic influence on the outcome[12, 13].

To study the dynamical evolution of conducting channel formation of our thin film structure, a self-consistent solution of Equation 6a and Equation 6b is obtained by the Multiphysics Object-Oriented Simulation Environment (MOOSE) finite-element platform[31, 32, 33, 34]. Periodic boundary conditions are imposed on the left and right edges of the discretized thin film structure of Figure 2 for both electric potential, $V(r, t)$, and concentration, $c(r,t)$. On the top and bottom edges Dirichlet boundary conditions for electric potential of 1.0 V and 0 V are imposed, respectively. Dirichlet boundary conditions for concentration are imposed on the top and bottom edges for an ideal electrical contact. The initial conducting and non-conducting regions are separated by an interface located at 4 nm from the bottom contact. The initial concentration, $c(r,0)$, is uniformly distributed between 0.1 and 0.3 for the non-conducting state and 0.7 and 0.9 for the conducting state, to approximate the pristine thinfilm structure shown in Figure 2. The double-well potential, $A$, is set to 1.0 eV ; the interfacial gradient energy term, $\kappa$, is set to 5.0 eV /nm$^2$ ; mobility, $M$ , is set for 0.1 nm$^2$ /(V × ns)[35, 36, 37, 38]. Convergence was defined by reaching a total energy minimum.

## III. DISCUSSION

An initial simulation was performed on an ideal abrupt interface to study the impact of interface roughness on the initial growth process of the conducting channel, as illustrated by the initial and final states illustrated by Figures 3a and 3b, respectively. The initial concentration of the conducting and non-conducting states of the structure were set uniformly to 0.7 and 0.3, respectively, thereby forming an abrupt interface with no roughness. As seen in Figure 3b, no conducting channel growth was observed, which suggests that the formation of conducting channels that bridge the top and bottom contacts is not energetically favored if the initial interface is abrupt. In other words, conducting channels may form if the initial interface has a certain level of roughness, which is consistent with experiments in which an interface separating two domains always have roughness. While the initial interface of Figure 3a appears to be morphologically abrupt, interface roughness in our modeling relates to the magnitude of the variation in $c(r,0)$ along the interface. An abrupt interface obtains when $c(r,0)$ is constant within the conducting and non-conducting states and a diffuse interface obtains when $c(r,0)$ varies within each of the two regions, as they do with initial conditions illustrated by Figure 2.

Using the pristine initial conditions specified by Figure 2, illustrated by Figure 3c, the self-consistent solution of Equations 6a and 6b for concentration, *c(r,t)*, is shown by Figure 3d. The results show formation of several conducting clusters and one continuous conducting channel, distinguished by the conducting domains shown in red and the non-conducting domains shown in blue. These results suggest unique interface microstructure develops from the initial interface under the influence of electrical potential and leading to the birth of distinctive conducting domains running through the non-conducting states. Note the presence of a continuous conducting channel traversing the bottom edge to the top edge of the thin film structure, as well as the presence of incomplete and orphaned conducting domains. The presence of a continuous conducting channel suggests the existence of an electroformed state, for the particular conditions of the present simulation.

The equilibrium interface formed by the conducting state and non-conducting state intrinsically describes the morphology of the conducting channel, represented by an envelope of cluster-like domains composed of many discrete charge carriers. The minimum energy condition imposed by the variation of the free-energy functional Equation 3 reflects local equilibrium between bulk free-energy density, interface energy, and their interaction with the applied electric potential, subject to appropriate boundary conditions and material properties.

To explore the influence of film thickness on the formation of conducting channels, an additional simulation was carried out on an otherwise identical thin film structure 50 nm thick, as shown by Figure 3e. Figure 3f shows the self-consistent solution for concentration, *c(r,t)*, under these new conditions. It is evident that while formation of several conducting channel-like clusters has occurred, there does not appear to be a continuous conducting channel traversing the bottom edge to the top edge of the thin film structure. In further contrast, there appears to be relatively more incomplete and orphaned conducting domains.

The thickness-to-width ratio for the structures of Figures 3c and 3e is 0.2 and 1.0, respectively. This suggests an affinity for conducting channel formation for low thickness-to-width ratios, perhaps because of the dominance of the bulk free-energy, as indicated by the free-energy functional Equation 3. This observation has also been previously suggested based on experimental observations the formation of conducting channels depends substantially on the relative aspect ratio of the thin film structure as well as a numerous experiments that seem to indicate that a dielectric thin film needs to be in the range of 1 nm to 10 nm for conducting channels to form[20].

## IV. SUMMARY

A computational phase field study of thin film conducting channel morphology and evolution has been presented. Atomic metastable states of thin films, responsible for resistance switching behavior, produces complex microstructure resulting from interaction of bulk free-energy and interface energy. The phase field formulation naturally avoids the mathematically onerous problem of tracking the dynamical evolution of the interface formed by this microstructure.

Our computational results suggest the phase field formulation can model the dynamical evolution of conducting channel formation and growth, illustrating a new method for the study of dielectric thin films for resistive switching. Our results further suggest that only when the initial interface has roughness do conducting channels form. Since interface roughness is expected to exist in any real dielectric thin film interface, we thus expect to observe switching in many films.

Furthermore, even though an initial well-prepared interface will exhibit finite roughness, the film thickness needs to be thin enough for a conducting channel to form, consistent with experimental data from thin film structures having a thickness-to-width ratio substantially less than unity. Physically this makes sense insofar as the lateral dimension establishes the number of nucleation sites at which conducting channels start forming, i.e. if the lateral dimension is extremely small, there exists reduced likelihood of conducting channel formation.

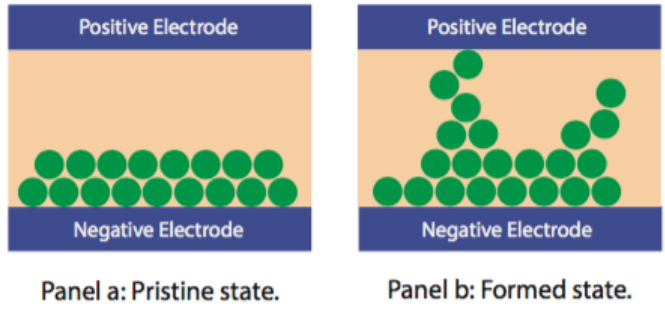

Panel a: Pristine state.  Panel b: Formed state.

FIG. 1. The conducting channel formalism illustrated by a charged species cluster model. Panel (a) illustrates the pristine pre-formed state; panel (b) illustrates a possible formed state, showing one complete conducting channel forming a continuous path between the negative and positive contacts. The green circles represent discrete charged species, hosted by a dielectric, shown in tan. The phase field formulation tracks the dynamical evolution of the envelope of clusters of these charged species, whose interface collectively constitutes a conductive channel, subject to a variational principle and local conservation laws.

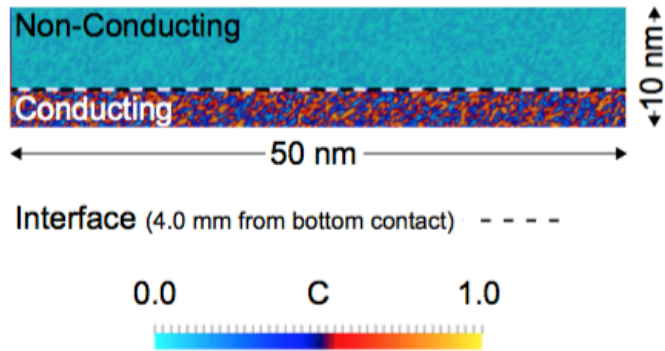

FIG. 2. A pristine dielectric thin film structure composed of a conducting region and a non-conducting region to approximate an as-fabricated resistive switching device. The interface is represented by the dotted horizontal line, and is 4 nm from the bottom edge. The pristine concentration is established by an initial concentration, $c(\vec{r}, 0)$, uniformly distributed between 0.1 and 0.3 for the non-conducting state and 0.7 and 0.9 for the conducting state. The structure is 50 nm × 10 nm.

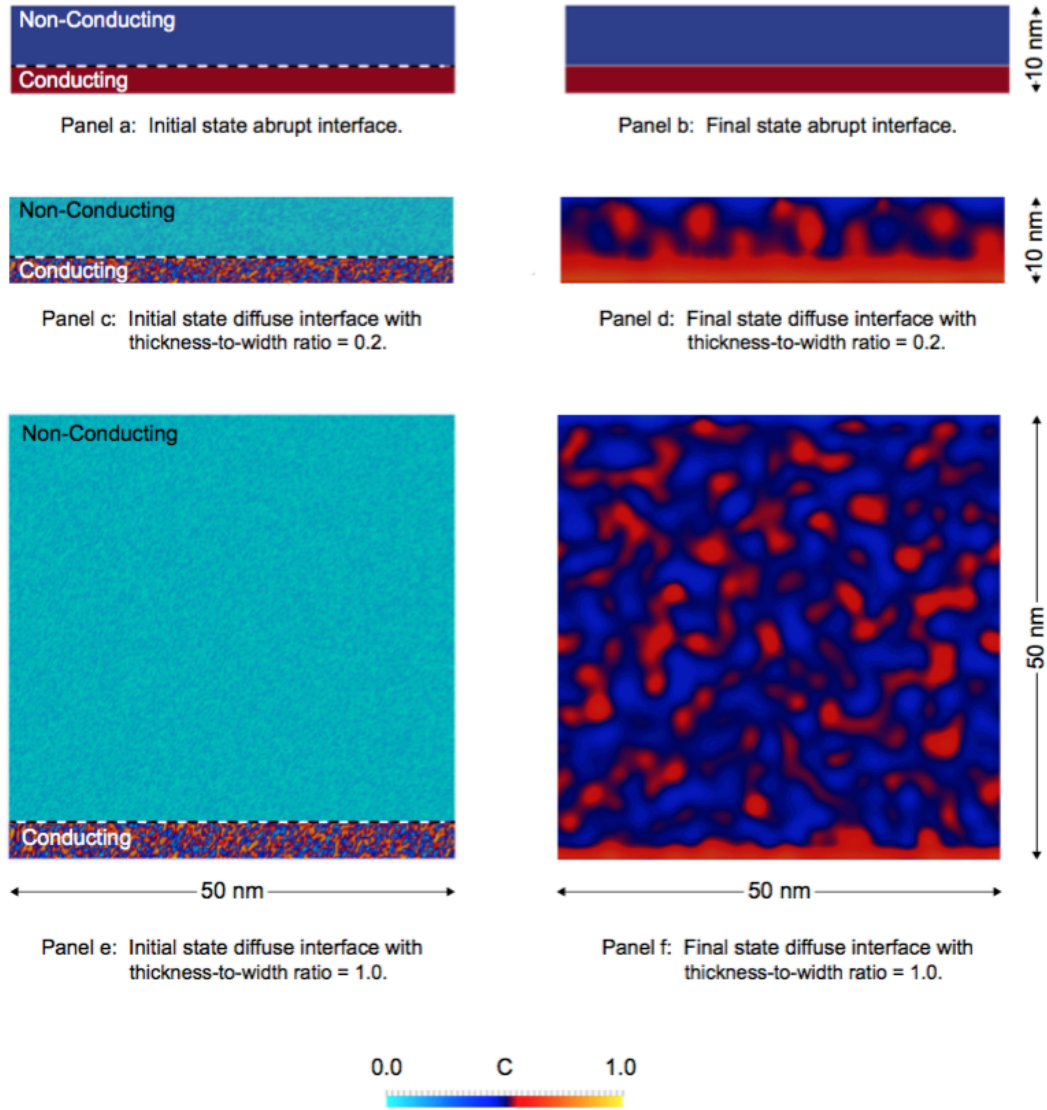

FIG. 3. Initial state and the final state self-consistent solution of Equations 6a and 6b for concentration, $c(\vec{r},t)$. The initial interface is formed 4 nm from the bottom contact. Panel (a) shows an ideal abrupt interface, defined as $c(\vec{r},0)$ being constant within the conducting and non-conducting regions, with panel (b) showing the resultant solution for $c(\vec{r},t)$. The initial conditions of this abrupt interface are established by uniformly setting the conducting and non-conducting concentrations of the structure to 0.7 and 0.3, respectively. In this case, no conducting channel growth was observed, suggesting formation of conducting channels that bridge the top and bottom contacts is not energetically favored if the initial interface is abrupt. Panel (c) shows a diffuse interface, established by an initial concentration, $c(\vec{r},0)$, uniformly distributed between 0.1 and 0.3 for the non-conducting state and 0.7 and 0.9 for the conducting state, reproducing the pristine thin-film structure shown in Figure 2. Panel (d) showing the resultant solution for $c(\vec{r},t)$, showing formation of several conducting clusters and one continuous conducting channel, distinguished by the conducting domains shown in red and the non-conducting domains shown in blue. Panel (e) applies the initial conditions of panel (c) to a 50 nm × 50 nm structure, with the resultant solution for $c(\vec{r},t)$ shown by panel (f). It is evident that while formation of several conducting channel-like clusters has occurred, there does not appear to be a continuous conducting channel traversing the bottom edge to the top edge of the thin film structure.

[1] L. D. Landau and E. M. Lifshitz, *Statistical Physics*, 3rd ed. (Pergamon Press, Elmsford, New York, 1980).

[2] N. F. Mott, "Metal-insulator transition," Rev. Mod. Phys. **40**, 677–683 (1968).

[3] M. Imada, A. Fujimori, and Y. Tokura, "Metal-insulator transitions," Rev. Mod. Phys. **70**, 1039–1263 (1998).

[4] R. Waser and M. Aono, "Nanoionics-based resistive switching memories," Nature materials **6**, 833–40 (2007).

[5] M. Seul and D. Andelman, "Domain shapes and patterns: The phenomenology of modulated phases," Science **267**, 476–483 (1995).

[6] Z. Hiroi, H. Hayamizu, T. Yoshida, Y. Muraoka, Y. Okamoto, J.-i. Yamaura, and Y. Ueda, "Spinodal decomposition in the tio2vo2 system," Chemistry of Materials **25**, 2202–2210 (2013).

[7] N. Xu, L. Liu, X. Sun, X. Liu, D. Han, Y. Wang, R. Han, J. Kang, and B. Yu, "Characteristics and mechanism of conduction/set process in tinznopt resistance switching random-access memories," Applied Physics Letters **92**, 232112 (2008).

[8] F. Pan, C. Chen, Z. shun Wang, Y. chao Yang, J. Yang, and F. Zeng, "Nonvolatile resistive switching memories-characteristics, mechanisms and challenges," Progress in Natural Science: Materials International **20**, 1 – 15 (2010).

[9] D. Ielmini, "Modeling the universal set/reset characteristics of bipolar rram by field- and temperature driven filament growth," IEEE Trans. Electron Devices **58**, 4309 (2011).

[10] T. Ahmed, S. Walia, E. L. Mayes, R. Ramanathan, P. Guagliardo, V. Bansal, M. Bhaskaran, J. J. Yang, and S. Sriram, "Inducing tunable switching behavior in a single memristor," Applied Materials Today **11**, 280 – 290 (2018).

[11] G. Gibson, S. Musunuru, J. Zhang, K. Vandenberghe, J. Lee, C. Hsieh, W. Jackson, Y. Jeon, Z. Li, and R. Williams, "An accurate locally active memristor model for s-type negative differential resistance in nbox," Applied Physics Letters **108**, 023505 (2016).

[12] J. F. Sevic and N. P. Kobayashi, "Multi-physics transient simulation of monolithic niobium dioxide-tantalum dioxide memristor selector structures," Applied Physics Letters **111**, 153107 (2017), https://doi.org/10.1063/1.5003168.

[13] J. F. Sevic and N. P. Kobayashi, "Self-consistent continuum-based transient simulation of electroformation of niobium oxide tantalum oxide selector-memristor structures," Journal of Applied Physics **124**, 164501 (2018).

[14] J. J. Yang, F. Miao, M. D. Pickett, D. A. A. Ohlberg, D. R. Stewart, C. N. Lau, and R. S. Williams, "The mchanism of electroforming of metal oxide memristive switches," Nanotechnology **20**, 215201 (2009).

[15] J. P. Strachan, M. D. Pickett, J. J. Yang, S. Aloni, A. L. David Kilcoyne, G. Medeiros-Ribeiro, and R. Stanley Williams, "Direct identification of the conducting channels in a functioning memristive device," Advanced Materials **22**, 3573–3577 (2010).

[16] F. Miao, J. P. Strachan, J. J. Yang, M.-X. Zhang, I. Goldfarb, A. C. Torrezan, P. Eschbach, R. D. Kelley, G. Medeiros-Ribeiro, and R. S. Williams, "Anatomy of a nanoscale conduction channel reveals the mechanism of a high-performance memristor," Advanced Materials **23**, 5633–5640 (2011).

[17] F. Nardi, S. Larentis, S. Balatti, D. Gilmer, and D. Ielmini, "Resistive switching by voltage-driven ion migration in bipolar rrampart i: Experimental study," IEEE Trans. Electron Devices Lett. **59**, 2461 (2012).

[18] S. Larentis, . F. Nardi, , S. Balatti, D. Gilmer, and D. Ielmini, "Resistive switching by voltage-driven ion migration in bipolar rram - part ii: Modeling," IEEE Trans. Electron Devices Lett. **59**, 2468 (2012).

[19] S. Kim, S. Kim, K. M. Kim, S. R. Lee, M. Chang, E. Cho, Y.-B. Kim, C. J. Kim, U. In Chung, and I.-K. Yoo, "Physical electro-thermal model of resistive switching in bi-layered resistance-change memory," Scientific Reports **3**, 1680 EP – (2013).

[20] D. B. Strukov, G. S. Snider, D. R. Stewart, and R. S. Williams, "The missing memristor found," Nature **453**, 80 EP – (2008).

[21] K. M. Kim, T. H. Park, and C. S. Hwang, "Dual conical conducting filament model in resistance switching tio2 thin films," Scientific Reports **5**, 7844 EP – (2015).

[22] W. J. Boettinger, J. A. Warren, C. Beckermann, and A. Karma, "Phase-field simulation of solidification," Annual Review of Materials Research **32**, 163–194 (2002).

[23] L.-Q. Chen, "Phase-field models for microstructure evolution," Annual Review of Materials Research **32**, 113–140 (2002).

[24] Q. C. Sherman and P. W. Voorhees, "Phase-field model of oxidation: Equilibrium," Phys. Rev. E **95**, 032801 (2017).

[25] W. Shen, N. Kumari, G. Gibson, Y. Jeon, D. Henze, S. Silverthorn, C. Bash, and S. Kumar, "Effect of annealing on structural changes and oxygen diffusion in amorphous hfo2 using classical molecular dynamics," Journal of Applied Physics **123**, 085113 (2018).

[26] N. Provatas and K. Elder, *Phase-Field Methods in Materials Science and Engineering* (Wiley-VCH Verlag, Weinheim, Germany, 2010).

[27] The normalization constant has been suppressed.

[28] For the current formulation, $\Omega$ is the volume of a mesh cell, of unit depth, following finite-element discretization of the structure of Figure 2.

[29] J. W. Cahn and J. E. Hilliard, "Free energy of a nonuniform system. i. interfacial free energy," The Journal of Chemical Physics **28**, 258–267 (1958).

[30] The functional spatial and time dependence of $V(\vec{r}, t)$ and $c(\vec{r}, t)$ has been dropped for Equation 6a.

[31] D. Gaston, C. Newman, G. Hansen, and D. Lebrun-Grandié, "Moose: A parallel computational framework for coupled systems of nonlinear equations," Nuclear Engineering and Design **239**, 1768–1778 (2009).

[32] B. S. Kirk, J. W. Peterson, R. H. Stogner, and G. F. Carey, "`libMesh`: A C++ Library for Parallel Adaptive Mesh Refinement/Coarsening Simulations," Engineering with Computers **22**, 237–254 (2006).

[33] S. Balay, S. Abhyankar, M. F. Adams, J. Brown, P. Brune, K. Buschelman, L. Dalcin, V. Eijkhout, W. D. Gropp, D. Kaushik, M. G. Knepley, L. C. McInnes, K. Rupp, B. F. Smith, S. Zampini, H. Zhang, and H. Zhang, "PETSc users manual," Tech. Rep. ANL-95/11 - Revision 3.7 (Argonne National Laboratory, 2016).

[34] M. R. Tonks, D. Gaston, P. C. Millett, D. Andrs, and P. Talbot, "An object-oriented finite element framework for multiphysics phase field simulations," Computational Materials Science **1**, 20 – 29 (2012).

[35] K. Stella, D. A. Kovacs, D. Diesing, W. Brezna, and J. Smoliner, "Charge transport through thin amorphous titanium and tantalum oxide layers," Journal of The Electrochemical Society **158**, P65–P74 (2011).

[36] D. B. Strukov and R. S. Williams, "Exponential ionic drift: fast switching and low volatility of thin-film memristors," Applied Physics A **94**, 515–519 (2009).

[37] U. N. Gries, H. Schraknepper, K. Skaja, F. Gunkel, S. Hoffmann-Eifert, R. Waser, and R. A. De Souza, "A sims study of cation and anion diffusion in tantalum oxide," Phys. Chem. Chem. Phys. **20**, 989–996 (2018).

[38] T. Reed, *Free Energy of Formation of Binary Compounds*, 1st ed. (MIT Press, Cambridge, MA, 1971).